\documentclass[pra,twocolumn,superscriptaddress,a4paper]{revtex4}
\usepackage{epsfig}
\usepackage{graphics}
\newcommand{\text}[1]{{\rm #1}}

\begin{document}

\title{Long-Range Order in Electronic Transport through Disordered Metal Films}
\author{S. Aigner}
 \affiliation{Atominstitut der \"Osterreichischen Universit\"aten, TU-Wien, Stadionallee 2, 1020 Vienna, Austria}
 \affiliation{Physikalisches Institut, Universit\"at Heidelberg, Philosophenweg 12, D-69120 Heidelberg, Germany}
\author{L. Della Pietra}
 \affiliation{Physikalisches Institut, Universit\"at Heidelberg, Philosophenweg 12, D-69120 Heidelberg, Germany}
\author{Y. Japha}
 \affiliation{Department of Physics, Ben-Gurion University of the Negev, P.O. Box 653, Be'er-Sheva 84105, Israel}
\author{O. Entin-Wohlman}
 \affiliation{Department of Physics, Ben-Gurion University of the Negev, P.O. Box 653, Be'er-Sheva 84105, Israel}
\author{T. David}
 \affiliation{Department of Physics, Ben-Gurion University of the Negev, P.O. Box 653, Be'er-Sheva 84105, Israel}
\author{R. Salem}
 \affiliation{Department of Physics, Ben-Gurion University of the Negev, P.O. Box 653, Be'er-Sheva 84105, Israel}
\author{R. Folman}
 \affiliation{Department of Physics, Ben-Gurion University of the Negev, P.O. Box 653, Be'er-Sheva 84105, Israel}
\author{J. Schmiedmayer}
 \affiliation{Atominstitut der \"Osterreichischen Universit\"aten, TU-Wien, Stadionallee 2, 1020 Vienna, Austria}
 \affiliation{Physikalisches Institut, Universit\"at Heidelberg, Philosophenweg 12, D-69120 Heidelberg, Germany}

\begin{abstract}

Ultracold atom magnetic field microscopy enables the probing of
current flow patterns in planar structures with unprecedented
sensitivity.  In polycrystalline metal (gold) films we observe
long-range correlations forming organized patterns oriented at
$\pm 45^\circ$ relative to the mean current flow, even at room
temperature and at length scales orders of magnitude larger than
the diffusion length or the grain size. The preference to form
patterns at these angles is a direct consequence of universal
scattering properties at defects. The observed amplitude of the
current direction fluctuations scales inversely to that expected
from the relative thickness variations, the grain size and the
defect concentration, all determined independently by standard
methods. This indicates that ultracold atom magnetometry enables
new insight into the interplay between disorder and transport.
\end{abstract}

\maketitle

%
%
\newpage

Thin metal films are the classic environment for studying the
effect of geometric constraints \cite{Ref:Sondheimer,Ref:Chambers}
and crystal defects \cite{Ref:MayadasShatzkes,Ref:Landauer} on the
transport of electrons. In a perfectly straight long wire that is
free from structural defects, a direct current (DC) strictly follows the
wire direction and creates a magnetic field in the plane
perpendicular to the wire. An obstacle may locally change the
direction of the current and consequently locally rotate the
magnetic field close to the wire by an angle $\beta$ in a plane
parallel to the plane of the thin film wire.

Ultracold atom magnetometry \cite{Ref:Wildermuth,WildermuthAPL} on
atom chips \cite{folman2002,ReichelReview,Fortagh2007} allows for
the sensitive probing of this angle $\beta$ (and its spatial
variation) with $\mu rad$ ($\mu m$) resolution. Compared to
scanning probes having a $\mu m$ scale spatial resolution and
$10^{-5}$T sensitivity, or superconducting quantum interference
devices (SQUIDs) having $10^{-13}$T sensitivity but a resolution
of tens of $\mu m$, ultracold atom magnetometry has both high
sensitivity ($10^{-10}$T) and high resolution (several $\mu m$)
\cite{WildermuthAPL}. In addition, ultracold atoms enable high
resolution over a large length scale ($mm$) in a single shot. This
enables the simultaneous observations of microscopic and
macroscopic phenomena, as described in this work.

Using cold atoms just above the transition to Bose-Einstein
Condensation (BEC), we apply ultracold atom magnetometry to study
the current deflection in three different precision-fabricated
polycrystalline gold wires with a rectangular cross section of
$200\mu$m width and different thicknesses and crystalline grain
sizes, as summarized in Table~\ref{Table1} \cite{WeissLab}.
Choosing the wire length along x, its width along y and thickness
along z, Fig. 1 shows the maps of the angular variations
$\beta(x,y,z_0)=\delta B_x(x,y,z_0)/B_y$ of the magnetic field
created by a current of 180 mA flowing along the wire, measured at
$z_0=$3.5$\mu$m above its center (far from the edges).

\begin{figure}[t]
 \hspace{5mm} \includegraphics[width= 0.9 \columnwidth]{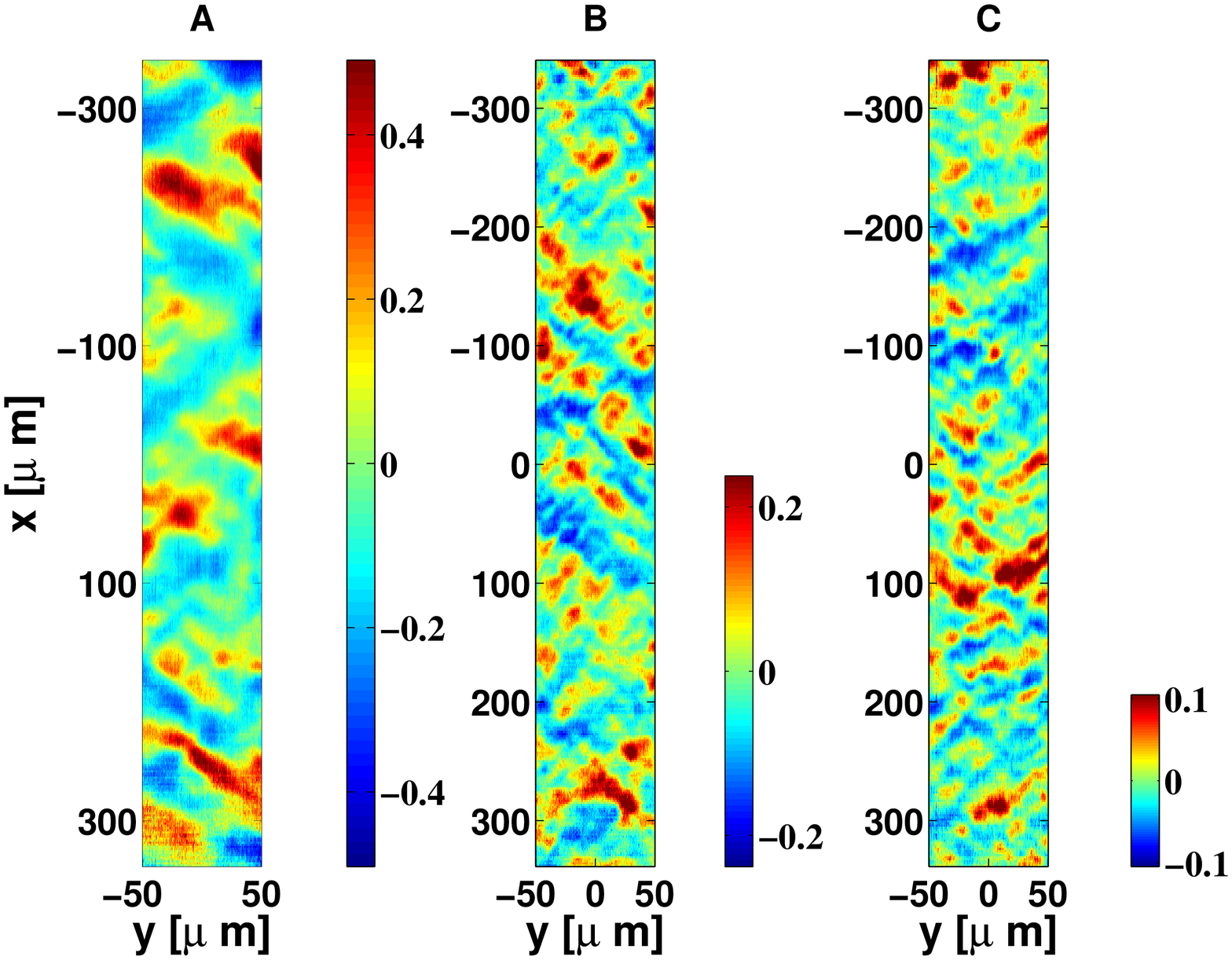} \\
 \vspace{3mm}
 \includegraphics[width= 0.9 \columnwidth]{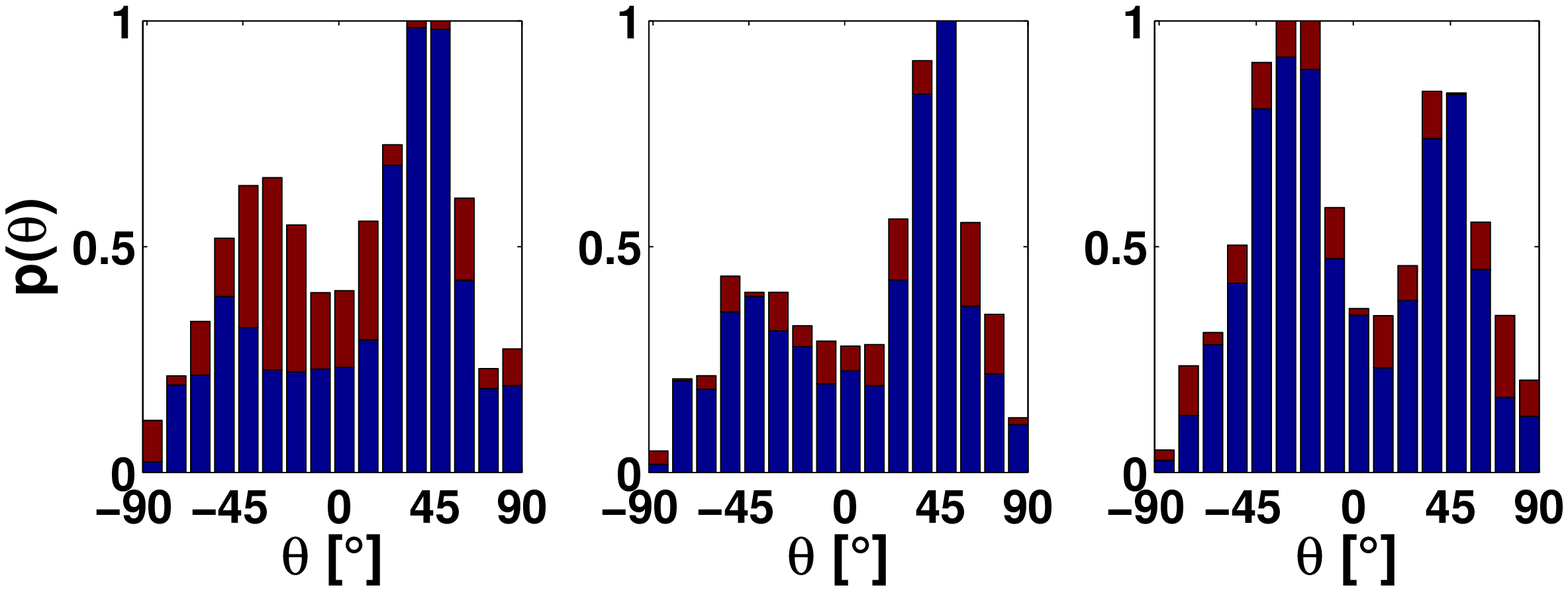}
\caption {(top) Magnetic field angle fluctuations $\beta
[\text{mrad}]$ above three (A-C) different polycrystalline gold
films described in Table~\protect\ref{Table1}. These fluctuations
are due to variations in the direction of the current flow and are
thus sensitive to $\delta J_y$. The appearance of $\pm 45^{\circ}$
patterns is clearly observable, and reflects a correlated
scattering of the electrons. (bottom) Quantification of the
angular pattern by the normalized angular power spectral density
$p(\theta)/\max(p)$. The red bars indicate the error.}
\end{figure}

Even though at ambient temperature scattering by lattice
vibrations (phonons) quickly diffuses the electronic motion,
long-range correlations (tens of $\mu m$) in the current flow
patterns can be seen. This is surprising as effects of static
defects are usually observed only on a length-scale of several
nanometers \cite{Ref:Schneider,Steinho"gel}. We observe clear
patterns of elongated regions of maximal current flow deviations
$\beta$ inclined by about $\pm 45^{\circ}$ to the mean current
flow direction. This angular preference is present in all the
measurements, independent of wire thickness or grain size. This
preference can be quantified by the normalized angular power
spectra $p(\theta)=\int dk k |\beta(k,\theta)|^2$ of the magnetic
field patterns, where ${\bf k}$ is the wavevector of the Fourier
transform of the measured $\beta(x,y)$ (Fig. 1).

\begin{table}[t]
\centering
\begin{tabular}{|l|c|c|c|} \hline
 wire & {\bf A} & {\bf B} & {\bf C} \\ \hline
 thickness $H$ ($\mu$m) & 2.08 & 0.28 & 0.28 \\ \hline
 grain size (nm) & 60-80 & 30-50 & 150-170 \\ \hline
resistivity at 296K ($\mu\Omega\cdot$cm) &2.73 & 3.1 & 2.77 \\
\hline resistivity at 4.2K ($\mu\Omega\cdot$cm) &0.094 & 0.316 &
0.351 \\ \hline atom temperature (nK) & $286 \pm 15$ & $173 \pm 2$
& $92 \pm 7$ \\ \hline
 measurement height ($\mu$m) & 3.5$\pm$0.4   & 3.4$\pm$0.3 & 3.7$\pm$0.4 \\ \hline
 $\delta z^{\rm rms}_+$ (AFM) (nm)& 9.4 & 3.5 & 3.1 \\ \hline
 $\delta z^{\rm rms}_+$ (WLI) (nm)& 1.31 & 0.42 & 0.48 \\ \hline
 $\delta z^{\rm rms}_+/H$ (WLI) ($\times 10^{-3}$) & 0.629 & 1.500 & 1.714 \\ \hline
 $\beta_{\rm rms}$ (mrad) &0.168 & 0.0715 & 0.0388 \\ \hline
 $\beta_{\rm pp}$ (mrad) &0.4 & 0.2 & 0.1 \\ \hline
 $\lambda_{\beta}$ ($\mu m$) &77 & 46 & 48 \\ \hline
\end{tabular}
\caption{Properties of the wires under investigation. All
measurements were done on the chip used for the cold atom
experiment except for the low temperature resistivity which was
measured on a duplicate chip made with an identical (simultaneous)
fabrication process. (AFM: Atomic force microscope; WLI: ZYGO
white light interferometer)} \label{Table1}
\end{table}

We observe significant difference in the magnitude and spectral
composition of the magnetic field variations above wires with
different thicknesses. Table 1 summarizes the main observations
and wire properties. The magnitude of $\beta$ scales contrary to
the surface corrugations when compared to the thickness; the
thinner films ($H=280$ nm) have the largest relative thickness
variations but show the smallest current directional variations.
Moreover, the thin wire with the large grains (grain size 150-170
nm) shows the smallest variations ($\beta_{\rm rms} = 39 \mu
rad$), much too small to be explained by the measured top surface
roughness $\delta z^{\rm rms}_+/H = 1.7 \times 10^{-3}$ of the
gold film.

The observed magnetic field variations caused by the current
direction variations are orders of magnitude smaller than the ones
reported in studies of 'fragmentation' of cold atom clouds on atom
chips (for a review see \cite{Fortagh2007}). These previously
reported fragmentation measurements can be fully explained by
corrugations in the wire edges \cite{lukin,orsay2}. In the present
study the effects caused by the wire edge roughness are strongly
suppressed by the much improved fabrication
\cite{fab_Groth,WeissLab}, and the flat wide wire geometry where
the ratio between the distance to the wire surface and to the wire
edge is very high \cite{krueger}.

%
%

In order to analyze the underlying mechanism for the current
direction deviations we consider a thin film (conductivity
$\sigma_0$) in the $x-y$ plane with a regular current $J_0 \hat{x}
=\sigma_0 {\bf E}^{(0)}$, where the electric field ${\bf E}^{(0)}$
is in the $\hat{x}$ direction. We consider the effect of small
fluctuations in the conductivity $\delta\sigma({\bf x})$ on the
current flow.

\begin{figure}[t]
 \includegraphics[width= 0.9 \columnwidth]{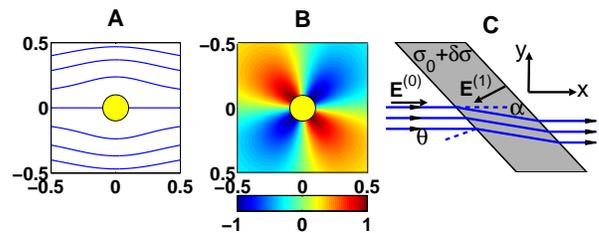}
 \caption{(A) Current scattering by circularly symmetric (disk)
local conductivity variations $\delta\sigma<0$. (B) The transverse
($\hat{y}$) component of the current is proportional to $\sin
2\theta$. (C) Direction change of a current flow due to a
conductivity step defect inclined by an angle $\theta$ relative to
the current flow direction $\hat{x}$. The conductivity is
$\sigma_0$ everywhere except in the shaded area, where it is
$\sigma_0 +\delta\sigma$ ($\delta\sigma>0$ in this example).
Again, the electron scattering amplitude is proportional to $\sin
2\theta$.}
\end{figure}

The current flow around a circular defect (Fig. 2A,B) generates a
dipole field with a transverse component $E_y^{(1)}\propto \sin
2\theta$, causing the current field to be repelled from (for
$\delta\sigma<0$) or attracted to (for $\delta \sigma>0$) the
defect and a $45^{\circ}$ pattern in the transverse current flow
forms.

A second illustration is a conductivity step ($\delta\sigma$)
inclined by an angle $\theta$ to the current flow direction (Fig.
2C). The resulting current density fluctuation is:
\begin{equation}
\delta {\bf J}= J_0\frac{\delta\sigma}{\sigma_0}
(\sin^2\theta\hat{x}-\cos\theta\sin\theta\hat{y}).
\label{eq:deltaJ}
\end{equation}
The transverse current component $J_y$ is again proportional to
$\sin 2\theta$, which is maximal for conductivity steps inclined
by $\theta=\pm 45^{\circ}$.

In a metal film, we expect to find a random pattern of
conductivity fluctuations $\delta\sigma({\bf x})$.  It can be
constructed from a random spatial distribution of the above basic
elements: microscopic circular defects or macroscopic conductivity
steps of different angles.

For a general quantitative analysis we expand an arbitrary
distribution $\delta\sigma({\bf x})$ in a Fourier series of plane
waves of the form $\delta \sigma({\bf x})= \delta\sigma_{\bf k}
\sin({\bf k}\cdot{\bf x}+\phi)$, where ${\bf
k}=(k_x,k_y)=k(\cos\theta_{\bf k},\sin\theta_{\bf k})$ and $\phi$
is an arbitrary phase. Each plane wave contributes to the current
fluctuation angle $\alpha=\delta J_y/J_0$ according to
Eq.~\ref{eq:deltaJ}, giving $\alpha({\bf k}) \approx -\sin
2\theta_{\bf k} (\delta\sigma_{\bf k}/2\sigma_0)$, and resulting
in the observed $45^{\circ}$ pattern.

The resulting magnetic field angle fluctuations at height $z$ above
the wire is directly related to the current fluctuations by
\begin{equation}
\beta({\bf k},z)\approx e^{-kz}\alpha({\bf k})\approx -\frac{1}{2}e^{-kz}
\frac{\delta\sigma_{\bf k}}{\sigma_0}\sin 2\theta_{\bf k},
\label{beta}
\end{equation}
which exhibits the same angular dependence. The exponential term
$e^{-kz}$ represents a resolution limit, such that the effect of
current changes on a length scale smaller than $2\pi z$ are
suppressed in the spectrum of the magnetic field fluctuations.
Starting from random conductivity fluctuations with a non-white
spatial frequency distribution the angular dependence $\sin
2\theta$ will emerge, giving rise to the observed $\pm 45^\circ$
preference. We have simulated such random models and the
observable $\beta$ forms very similar 2D maps as in Fig. 1.

%
%

The variations $\delta\sigma({\bf x})$ in the conductivity
$\sigma$ in a thin metal film are caused by contributions from two
physical origins: bulk conductivity variations in the metal, and
variations in the boundaries, namely variations in the thickness
$H$ of the film $\delta H({\bf x})$ leading to a change in the
conductivity per unit area $\delta\sigma=\sigma_0\delta H/H$.

In order to investigate whether the observed current flow
deviations are related to corrugations in the top surface of the
wire, we have measured the surface topography of the wires using a
white-light interferometer. No angular preference inherent in the
structure of the wires was found. Consequently, the angular
pattern in the magnetic field variations presented in Fig. 1 must
be a pure property of the scattering mechanism of the current flow
by the wire defects, as outlined above. More so, when we calculate
the two dimensional magnetic field at $3.5\mu m$ above the
surface, using the white-light interferometry measurements and the
assumption $\delta H({\bf x})=\delta z_+({\bf x})$, we could not
find a reasonable fit between the latter and the magnetic mapping
done by the atoms (Fig. 1). A detailed analysis of the top surface
corrugations $\delta z_+$ (Fig. 3) shows that they are
significantly larger for the thick film compared to the two thin
films, especially at short length scales.

\begin{figure}
 \includegraphics[width= 0.75 \columnwidth]{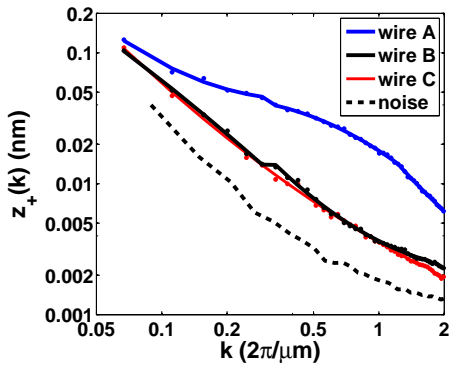}
 \caption{Radial spectrum of the top surface corrugations
$|\delta z_+(k)|=\sqrt{(2\pi)^{-1}\int d\theta |\delta
z_+(k,\theta)|^2}$ for the three wires measured using a
white-light interferometer. Note the significant difference
between wire \textbf{A} and wires \textbf{B}, \textbf{C}. (dashed)
Measurement noise level calculated by averaging over many partly
overlapping images}
\end{figure}

To quantify our findings, we compare the power spectra of the
measured magnetic field variations to those calculated from
several models based on the measured top surface variations, an
assumed bottom surface roughness, and possible inhomogeneities in
the bulk conductivity (Fig. 4).

We start with the two thin wires: \textbf{B} and \textbf{C}. The
measured power spectra of the magnetic field variations are
significantly lower (by two orders of magnitude for large
wavelengths) than predictions based on a model with a flat bottom
surface ($\delta H=\delta z_+$). If we assume that the top surface
exactly follows the bottom surface ($\delta z_+=\delta z_-$), a
lower bound on the influence of the surface on magnetic field
fluctuations can be obtained, as this configuration produces
vertical currents whose contribution to the longitudinal magnetic
field, to which our experiment is sensitive, is very small. The
measured data is in between these two cases.

A fair fit of the measured spectrum for the thin wires is obtained
if we assume that the top surface partially follows the
large-wavelengths fluctuations of the bottom surface while
independent fluctuations of the top surface exist in the shorter
scale.  For such a model we assume $\delta z_-(k)\approx\delta
z_+(k)e^{-(k/k_0)^2}$. Note that the resulting average thickness
variations are extremely small $|\delta H^{\rm rms}|=|\delta
z^{\rm rms}_+-\delta z^{\rm rms}_-| < 1 \AA$. This value of
$\delta H^{\rm rms}$ refers to length scales longer than $1\mu m$,
while much larger surface variations were measured on the scale of
the grains using the AFM (see Table~\ref{Table1}).

The situation is different for the thick wire \textbf{A} ($H=2 \mu
m$). Models assuming a flat bottom surface ($\delta H=\delta z_+$)
and models assuming a corrugated bottom surface $\delta z_-$ with
a spectrum similar to that of wire \textbf{B} and no correlations
with the top surface, both underestimate the measured magnetic
field variations.

\begin{figure}[t]
 \includegraphics[width= 0.95 \columnwidth]{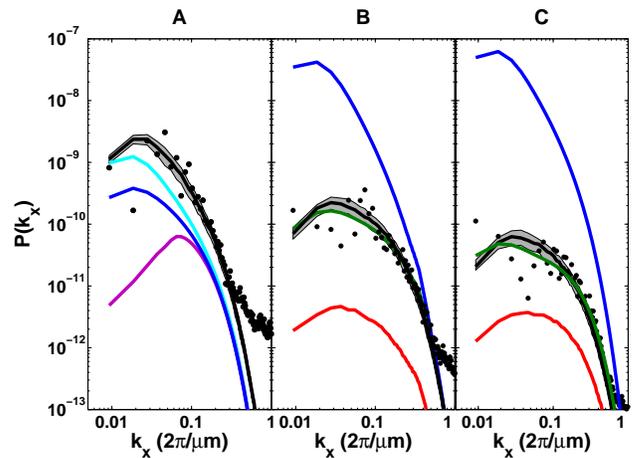}
 \caption{Comparison of surface and bulk model
calculations (lines) with the measured power spectrum
$P(k_x)=\sum_{k_y}|\beta(k_x,k_y)|^2$ of the magnetic field angle
$\beta$ along the $x$ direction above the three wires (points).
(blue) top surface $\delta z_+$ as in Fig. 3 with flat bottom
surface $\delta z_-=0$. (red) top surface follows bottom surface
$\delta z_+=\delta z_-$ (i.e. no thickness variations). (green)
Partially correlated top and bottom surfaces for wires \textbf{B}
and \textbf{C}. For the thick wire \textbf{A} we assume $\delta
z_-(k)$ as in wire \textbf{B}, which is correlated (purple) or
uncorrelated (light blue) with the top surface. The latter gives
the closest estimate for the experimental data, but gives
$\beta_{rms}$ which is only about half of the measured value.
(black) a fit to a model assuming bulk conductivity fluctuations.
The shaded area represents a one-standard-deviation range obtained
by varying the relative phases of different spectral components
$\delta\sigma(k_x,k_y)$.}
\end{figure}

The difference between the surface models and the measured data of
wire \textbf{A} can be attributed to fluctuations in the bulk
conductivity. A model taking the maximal contribution of surface
roughness (uncorrelated top and bottom surfaces) into account
gives the minimal required contribution of the bulk conductivity
fluctuations. If we apply the same minimal bulk conductivity
fluctuations as obtained from wire \textbf{A} to the two thin
wires \textbf{B} and \textbf{C}, they overestimate the measured
magnetic field fluctuations substantially for both wires and give
a different spectral shape. This indicates that the bulk
conductivity of the thinner wires should be more homogeneous than
that of the thick wire.

A more homogeneous bulk conductivity in the thin wires, however,
appears to be contradictory to the fact that the low temperature
resistivity is smaller for the thick wire than for the thin wires
(Table 1). Nevertheless, we note that this resistivity is mainly
determined by the small-scale properties of the wire (on the order
of the grain size or less) and by surface scattering, while the
magnetic field variations probe the conductivity inhomogeneities
at a larger scale and provide complementary information that would
not be available by standard methods.

Our analysis furthermore suggests that the differences in the
length scale $\lambda_{\beta}$ of the variation in $\beta$ as seen
in Fig. 1 and quantified in Table 1, may originate from the fact
that conductivity variations in the thin wires coming from
thickness variations are suppressed at long length scales due to
top and bottom surface correlations. In contrast, conductivity
variations in the thick wire originate at all scales from a
combination of thickness variations due to uncorrelated top and
bottom surfaces and a dominant contribution of bulk conductivity
inhomogeneity.

Our study is the first direct application of ultracold atoms as a
probe for solid state science. The exceptional sensitivity of the
ultracold atom magnetic field microscope
\cite{Ref:Wildermuth,WildermuthAPL} allowed us to observe
long-range patterns of the current flow in a disordered metal
film. The preference of features with angles around $\pm
45^{\circ}$ in the measured angular spectrum of the current flow
fluctuations is due to the universal scattering properties at
defects.  A detailed quantitative analysis shows that the observed
current directional fluctuations at different wires exhibits
significantly different and unexpected properties due to different
physical origins. This clearly demonstrates the power of the
ultracold atom magnetic field microscope to study details of the
current flow in conductors, and its ability to unveil previously
unaccessible information. This may be expected to stimulate new
studies on the interplay between disorder and coherent transport
in a variety of systems ranging from high-T$_c$ superconductors
\cite{Bonn} to 2D electron gases \cite{Ilani} and nano-wires
\cite{Feist}.

We thank the team of the Ben-Gurion University Weiss Family
Laboratory for Nanoscale Systems for the fabrication of the chip
and J\"urgen Jopp of the Ben-Gurion University Ilse Katz Center
for Nanoscale Science for assisting with surface measurements.
R.F. thanks Yoseph (Joe) Imry and Sir Aaron Klug for their
continued support. We acknowledge the support by the FWF, the DFG,
the German Federal Ministry of Education and Research (DIP), the
European Commission through the Integrated Project FET/QIPC
"SCALA" and the RTN "AtomChips", the American-Israeli Foundation
(BSF) and the Israeli Science Foundation.


\begin{thebibliography}{999}

\bibitem{Ref:Sondheimer}
    E. H. Sondheimer, {\it Adv. Phys.} {\bf 1}, 1 (1952).

\bibitem{Ref:Chambers}
    R. G. Chambers, {\it Proc. R. Soc. London, Ser. A} {\bf 202}, 378 (1950).

\bibitem{Ref:MayadasShatzkes}
    A. F. Mayadas, M. Shatzkes, {\it Phys. Rev. B} {\bf 1}, 1382 (1970).

\bibitem{Ref:Landauer}
    R. Landauer, {\it IBM J.Res. Dev.} {\bf 1}, 223 (1957).

\bibitem{Ref:Wildermuth}
    S. Wildermuth {\it et al.}, {\it Nature} {\bf 435}, 440 (2005).

\bibitem{WildermuthAPL}
    S. Wildermuth {\it et al.}, {\it Appl. Phys. Lett.} {\bf 88}, 264103 (2006).

\bibitem{folman2002}
    R. Folman, P.\ Kr\"uger, J.\ Schmiedmayer, J.\ Denschlag, C. Henkel, {\it Adv. At. Mol. Opt. Phys.} {\bf 48}, 263 (2002).

\bibitem{ReichelReview}
        J. Reichel, {\it Appl. Phys. B} {\bf 74}, 469 (2002)

\bibitem{Fortagh2007}
    J. Fortagh, J. C. Zimmermann, {\it Rev. Mod. Phys.} {\bf 79}, 235 (2007).

\bibitem{WeissLab} The chip was fabricated at The Weiss Family Laboratory for Nano-Scale Systems at Ben-Gurion University, Israel,    www.bgu.ac.il/nanofabrication.

\bibitem{Steinho"gel}
    W. Steinh\"ogl, G. Schindler, G. Steinlesberger, M. Engelhardt, {\it Phys. Rev. B} {\bf 66}, 075414 (2002).

\bibitem{Ref:Schneider}
    M. A. Schneider, M. Wenderoth, A. J. Heinrich, M. A. Rosentreter, R. G. Ulbrich, {\it Appl. Phys. Lett.} {\bf 69}, 1327 (1996).

\bibitem{lukin}
    D.-W.\ Wang, M.\ D.\ Lukin, E.\ Demler, {\it Phys. Rev. Lett.} {\bf 92}, 076802 (2004).

\bibitem{orsay2}
    T.\ Schumm {\it et al.}, {\it Euro. Phys. J. D} {\bf 32}, 171 (2005).

\bibitem{fab_Groth}
    S.\ Groth {\it et al.}, {\it Appl. Phys. Lett} {\bf 85}, 2980 (2004).

\bibitem{krueger}
    P.\ Kr\"uger {\it et al.}, arXiv:cond-mat/0504686; P.\ Kr\"uger {\it et al.}, \textit{Phys. Rev. A} \textbf{76}, 063621 (2007)

\bibitem{Bonn}
    D. A. Bonn, {\it Nature Physics} {\bf 2}, 159 (2006).

\bibitem{Ilani}
    S. Ilani, A. Yacoby, D. Mahalu,H. Shtrikman, {\it Science} {\bf 292}, 1354 (2001).

\bibitem{Feist}
    J. Feist {\it et al.}, {\it Phys. Rev. Lett.} {\bf 97}, 116804 (2006).


\end{thebibliography}
\end{document}